\documentclass{llncs}
\usepackage{makeidx}  
\usepackage{graphicx}
\begin{document}
\mainmatter              
\title{Automatic Right Ventricle Segmentation using Multi-Label Fusion in Cardiac MRI}
\titlerunning{Automatic RV Segmentation using Multi-Label Fusion in Cardiac MRI}  
%
\author{Maria A. Zuluaga
 \and M. Jorge Cardoso
  \and S\'{e}bastien Ourselin}

\authorrunning{Maria A. Zuluaga et al.} 
\institute{Centre for Medical Image Computing\\
University College London, London UK}

\maketitle              

\begin{abstract}
Accurate segmentation of the right ventricle (RV) is a crucial step in the assessment of the ventricular structure and function. Yet, due to its complex anatomy and motion segmentation of the RV has not been as largely studied as the left ventricle.  This paper presents a fully automatic method for the segmentation of the RV in cardiac magnetic resonance images (MRI). The method uses a coarse-to-fine segmentation strategy in combination with a multi-atlas propagation segmentation framework. Based on a cross correlation metric, our method selects the best atlases for propagation allowing the refinement of the segmentation at each iteration of the propagation. The proposed method was evaluated on 32 cardiac MRI datasets provided by the RV Segmentation Challenge in Cardiac MRI.

\end{abstract}
\section{Introduction}
An estimated 17.5 million people died from cardiovascular diseases in 2005, accounting 30\% of deaths around the world~\cite{who}. Over the last fifteen years, magnetic resonance imaging (MRI)  has become a routine modality for the determination of patient cardiac morphology. An accurate  extraction of the anatomical information is crucial to obtain reproducible quantitative measurements to support the diagnosis and follow-up of cardiac pathologies as well as to develop new clinical applications. Manual delineation of the anatomical structures is a long and tedious task, which is prone to inter- and intraobserver variations. Therefore, it is highly desirable to develop techniques for automatic segmentation.     

For years, left ventricle (LV) segmentation in cardiac MRI has been the main focus of several research groups. For instance, it has been the subject of two different segmentation challenges\footnote{MICCAI 2009 - Cardiac MRI Left Ventricle Segmentation Challenge and MICCAI 2011 - The STACOM'11 Cardiac Left Ventricular Segmentation Challenge}. We refer the reader to \cite{suri,petitjean2011} for detailed reviews on the topic. Due to its larger anatomical complexity, right ventricle (RV) segmentation has not been as extensively explored. However, recent work~\cite{caudron} has highlighted the importance of RV structure and function assessment in the management of cardiac diagnosis and the determinant role of a high quality RV segmentation.  The aim of this paper is to participate in the RV Segmentation Challenge in Cardiac MRI, an initiative to gather up and review state of the art methods in the field.

\begin{figure} 
\begin{center}
\begin{tabular}{ccc}
\includegraphics[scale=0.15]{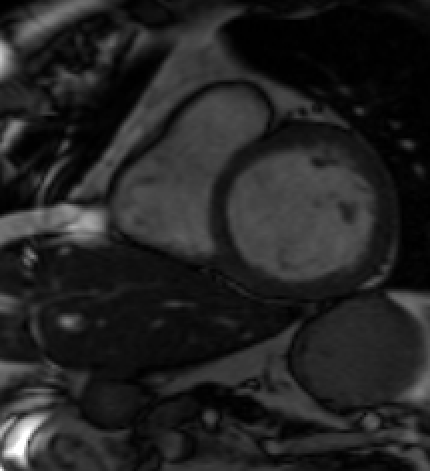} & \includegraphics[scale=0.15]{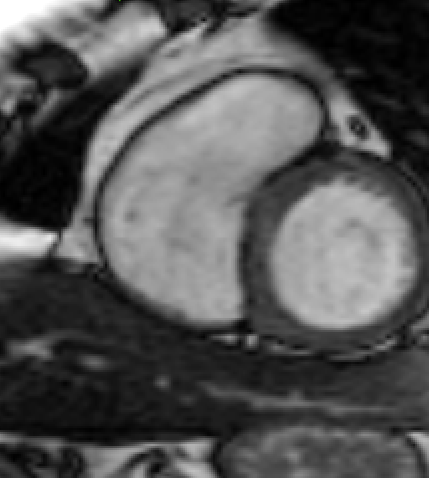} &  
\includegraphics[scale=0.15]{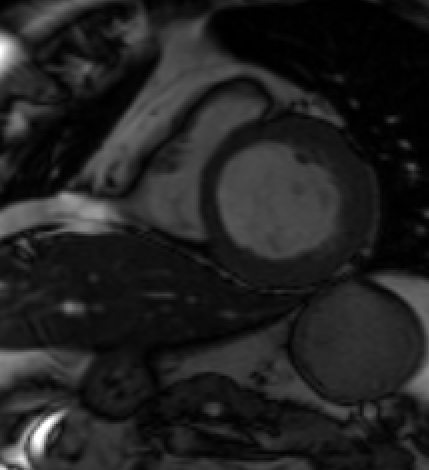} \\
\includegraphics[scale=0.15]{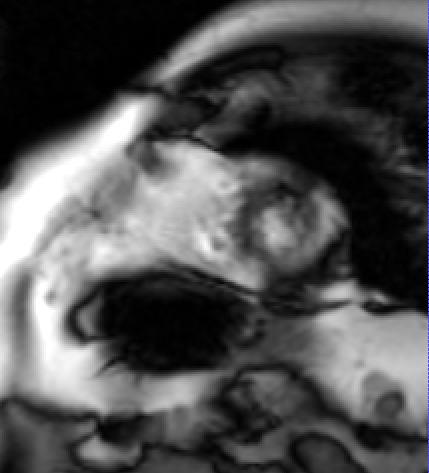} & 
\includegraphics[scale=0.15]{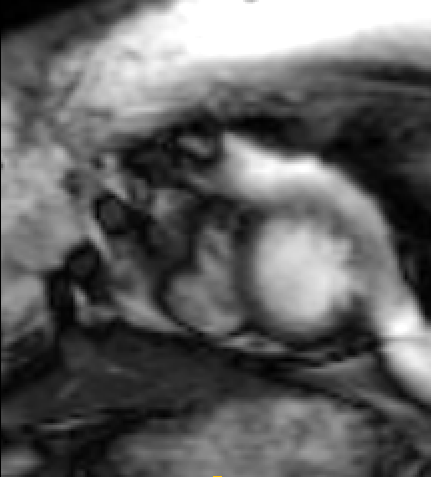} &
\includegraphics[scale=0.15]{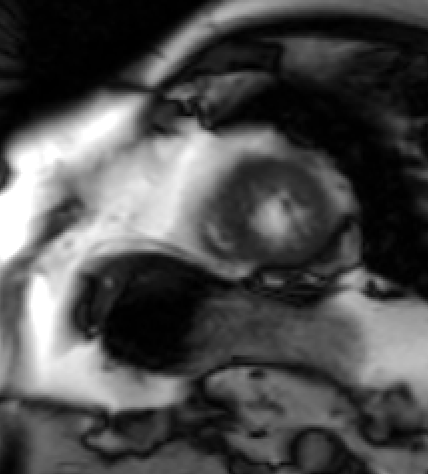} 
\end{tabular}
\end{center}
\caption{The shape of the ventricles can vary significantly between basal (top row) and apical slices (bottom row), end-diastole (left column) and systole (center column) and different patients (left and center column vs. right column). The use of atlases exhibiting a great variability can have a negative effect on the label fusion process.  } 
\label{fig:examples}
\end{figure}

In the remaining sections of the paper, we first describe the proposed segmentation approach, followed by the results obtained with the segmentation challenge data. Finally, we discuss on the obtained results and some conclusions are presented. 

\section{Methodology}
The RV segmentation method presented here is inspired on a previous approach for 3D whole heart segmentation using an atlas-based propagation segmentation propagation framework~\cite{zhuang} using a single atlas, combining an intensity image and a label image. Following this idea, we make use of multiple atlases for segmentation propagation in order to extract the endocardium and epicardium of the RV. 

\begin{figure} 
\begin{center}
\includegraphics[width=\columnwidth]{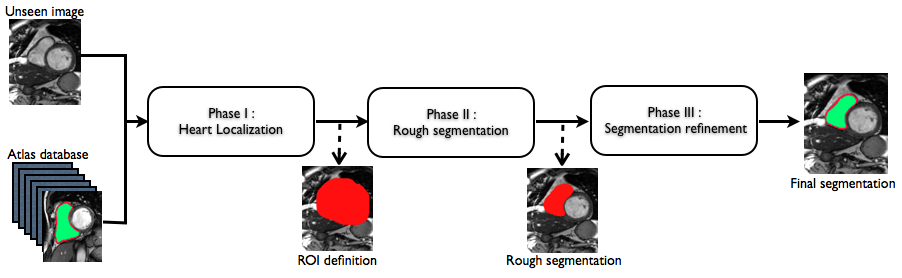} 
\end{center}
\caption{Automatic RV coarse-to-fine segmentation framework.  } 
\label{fig:framework}
\end{figure}

Multi-atlas segmentation propagation is typically performed  in two stages. First,  the unseen image is registered to all the intensity images of the atlases, followed by the mapping of the label images to the unseen image space using the obtained transformations. Then, at the second stage, the transformed labeled images are combined into a single segmentation of the unseen image through a label fusion method.

As our segmentation is performed on 2D and cardiac images can show large variability (Fig.~\ref{fig:examples}), it is necessary to perform an atlas selection that chooses the best suited atlases for a particular unseen image slice. For this matter, we use a multi-label fusion strategy based on the STAPLE algorithm
that incorporates a fast locally normalised cross correlation (LNCC) based ranking combined with 
a consensus based region-of-interest selection and an iterative Markov Random Field into the a multi-label STAPLE algorithm~\cite{warfield}. For further details on this multi-label fusion strategy, here denoted as STEPS, we refer the reader to~\cite{cardoso}. 

Instead of applying only once the multi-atlas propagation framework to an unseen image, we use a coarse-to-fine strategy that incrementally refines the segmentation by means of the multi-atlas propagation. The coarser segmentation obtained at each propagation level is used as a mask to start the registration gradually up to the finest level. The different steps of our coarse-to-fine strategy, illustrated in Fig.~\ref{fig:framework}, are described in the following. 

\paragraph{Phase I. Heart localization.} At a first stage, we seek to define a region of interest (ROI) that encloses the heart in the unseen image. At this phase, the unseen image is globally registered to the atlases using a block matching approach~\cite{ourselin}. The obtained transformations  are applied to the atlas' label images, which are all fused using  majority voting. Figure~\ref{fig:framework} shows the typical output ROI for the whole heart.

\paragraph{Phase II. Rough segmentation.} The unseen image ROI obtained from phase I is used as a mask in the registration step of phase II. The use of a mask allows the removal of structures surrounding the heart, e.g. the liver and the ribs, that can affect the registration procedure. 

At this phase, two different registration steps are perfomed. The atlas are rigidly registered~\cite{ourselin} to the masked unseen image, followed by a non-rigid alignment  using a fast free form deformation algorithm~\cite{modat}. After non-rigidly transforming the atlases, i.e. both intensity and label images, to the unseen image space, we apply an NCC-based ranking strategy~\cite{cardoso} in order to determine for each unseen slice  which are the most suitable registered  atlases slices to use. We selected only the 10\% best ranked atlases as several of the registrations fail to provide a good alignement of the unseen image and the atlases. The selected atlases slices were then fused using the STEPS algorithm~\cite{cardoso} in order to obtain an initial RV rough segmentation (Fig.~\ref{fig:framework}). As at this stage only the top 10\% atlases are used, the results from thie fusion step still don't have a high quality.

\paragraph{Phase III. Segmentation refinement.} In order to accurately initialize an affine  registration for all the atlases, all the label images are affinely aligned~\cite{ourselin} to the estimated rough segmentation. The transformed label images are fused through majority voting. As in phase II, this mask is used to remove surrounding structures in the final image-to-image non-rigid registration. With the non-rigid transformations, the label images are propagated and resampled using a nearest neighbor interpolation to preserve the binary nature of the segmentations. Finally, we make use of the multi-label fusion algorithm~\cite{cardoso} that first ranks every atlas slice based on the NCC and then fuses locally the labels of the N best ranked atlases $\left(N=\frac{\left|{atlas}\right|}{3}\right)$.

\begin{table} 
\begin{center}
\caption{Summary of the registration and fusion methods involved at each phase.} 
\label{tab:coarsetofinesummary} 
\begin{tabular}{llllll} 
\hline\noalign{\smallskip} 
\multicolumn{1}{c}{Phase} & \quad & \multicolumn{1}{c}{Registration} & \quad & \quad & \multicolumn{1}{c}{Fusion} \\ 
\noalign{\smallskip} 
\hline 
\noalign{\smallskip} 
I & & Image-to-image global rigid registration & & \quad & Majority Voting  \\ 
& & & &\\
II & & 1) Image-to-image affine registration  & & \quad & STEPS  \\
   & & 2) Image-to-image non-rigid registration & & \quad & \\ 
   & & & &\\
III & & 1) Mask-to-mask affine registration & & \quad & 1) Majority voting \\
    & & 2) Image-to-image non-rigid registration & & \quad & 2) STEPS \\
\hline 
\end{tabular}
\end{center}
\end{table} 

Table~\ref{tab:coarsetofinesummary} summarizes the different registration and fusion algorithms involved at each phase of the coarse-to-fine strategy.

\section{Experiments and Results}
The RV Segmentation Challenge in Cardiac MRI provided the participants with an initial set of 16 MR images containing manual annotations of the RV epicardium and endocardium 2D contours. We used these annotations to build up label images. These, in combination with the intensity MR images, made up our atlas set. As each dataset contained end-diastole (ED) and end-systole slices, separate atlases were built for ED and ES.  
\begin{table} 
\begin{center}
\caption{Image-based Dice metric (DM) and Hausforff distance (HD) of RV segmentation for the test data set.} 
\label{tab:results} 
\begin{tabular}{lllll} 
\hline\noalign{\smallskip} 
 & \quad  & \multicolumn{1}{c}{DM}  & \quad &\multicolumn{1}{c}{HD}  \\ 
 & \quad & \multicolumn{1}{c}{mean(std)} & \quad & \multicolumn{1}{c}{mean(std)}  \\
\noalign{\smallskip} 
\hline 
\noalign{\smallskip} 
\textbf{ED} & & & \\
Endocardium & & 0.83 (0.17) & &9.77 (7.88)  \\ 
Epicardium & &0.86 (0.13) & &10.23 (7.22) \\ 
\textbf{ES} & & & &\\
Endocardium & & 0.72 (0.27)& & 11.41 (10.49)  \\ 
Epicardium & & 0.77 (0.23) & &11.81 (9.46) \\
\hline 
\end{tabular}
\end{center}
\end{table} 

Training and parameter optimization of the algorithm was performed using a leave-one-out strategy using the training datasets. On a subsequent stage, the challenge organizers provided 16 additional cases (test set) to which our method was applied. In such scenario, the complete training set was used as an atlas. In average, the segmentation of a complete case (i.e. segmentation of both ED and ES) took 12 minutes. Evaluation of the results was performed in terms of both image- and patient-based criteria. Image-based criteria included the Dice metric (DM) and the Hausdorff distace (HD), whereas patient-based criteria, assessed endocardial volume at ED and ES, ejection fraction (EF) and ventricular mass (VM). Further details on the evaluation metrics can be found on the challenge website\footnote{http://www.litislab.eu/rvsc/}.

\begin{figure} 
\begin{center}
\begin{tabular}{ccc}
\includegraphics[width=0.2\columnwidth]{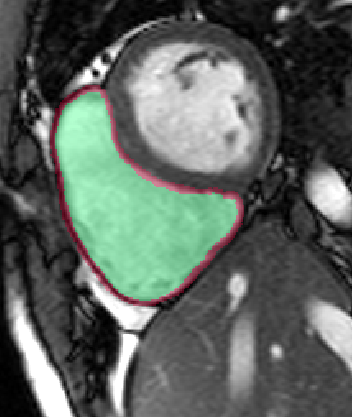} & \includegraphics[width=0.2\columnwidth]{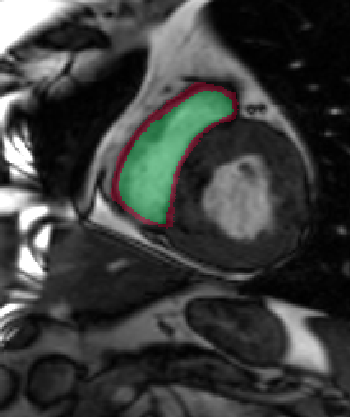} &  
\includegraphics[width=0.204\columnwidth]{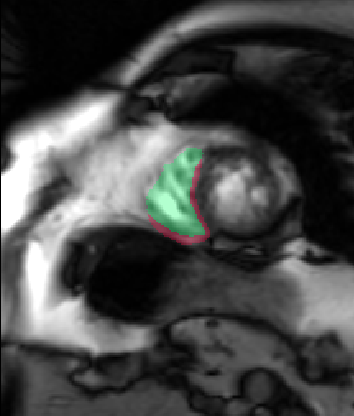} \\
\includegraphics[width=0.2\columnwidth]{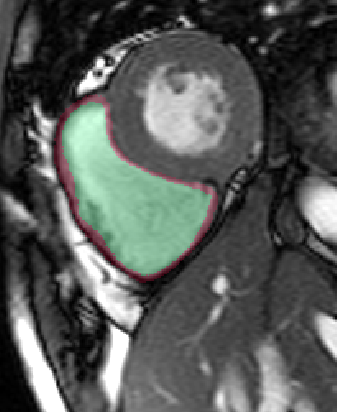} & 
\includegraphics[width=0.201\columnwidth]{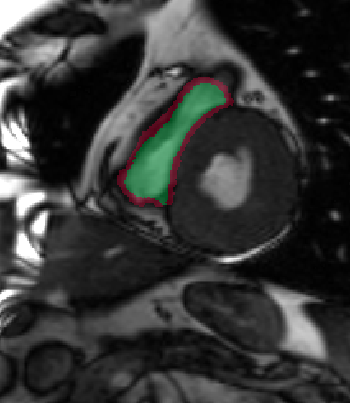} &
\includegraphics[width=0.201\columnwidth]{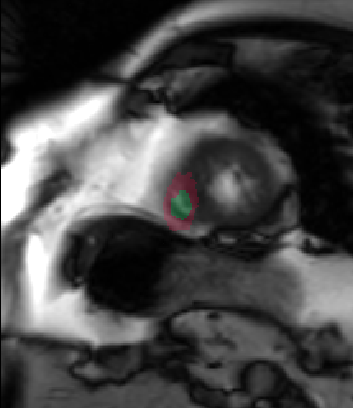} 
\end{tabular}
\end{center}
\caption{Example segmentations on cases 21 (left), 22 (center) and 01 (right) at both ED (top row) and ES (bottom row). Basal slices achieve high quality segmentations, whereas it fails  on apical slices, specially, in the epicardium extraction.} 
\label{fig:segmentations}
\end{figure}

Table~\ref{tab:results} summarizes the results of our method for RV segmentation in terms of image-based metrics. The results show that the algorithm performs better at ED than ES. This can be explained by the fact that the image quality is higher at ED than at ES. By analyzing the DM and HD on a slice-basis, it can be seen that the proposed method has a high performance on basal slices, with an average $DM=0.93$ and $HD=4.31$. However, the segmentations on apical slices are of lower quality affecting the overall DM and HD scores.  Figure~\ref{fig:segmentations} shows segmentation results in six different cases obtained from the challenge that illustrate the differences between basal and apical segmentations.

\begin{figure} 
\begin{center}
\begin{tabular}{cc}
\includegraphics[width=0.45\columnwidth]{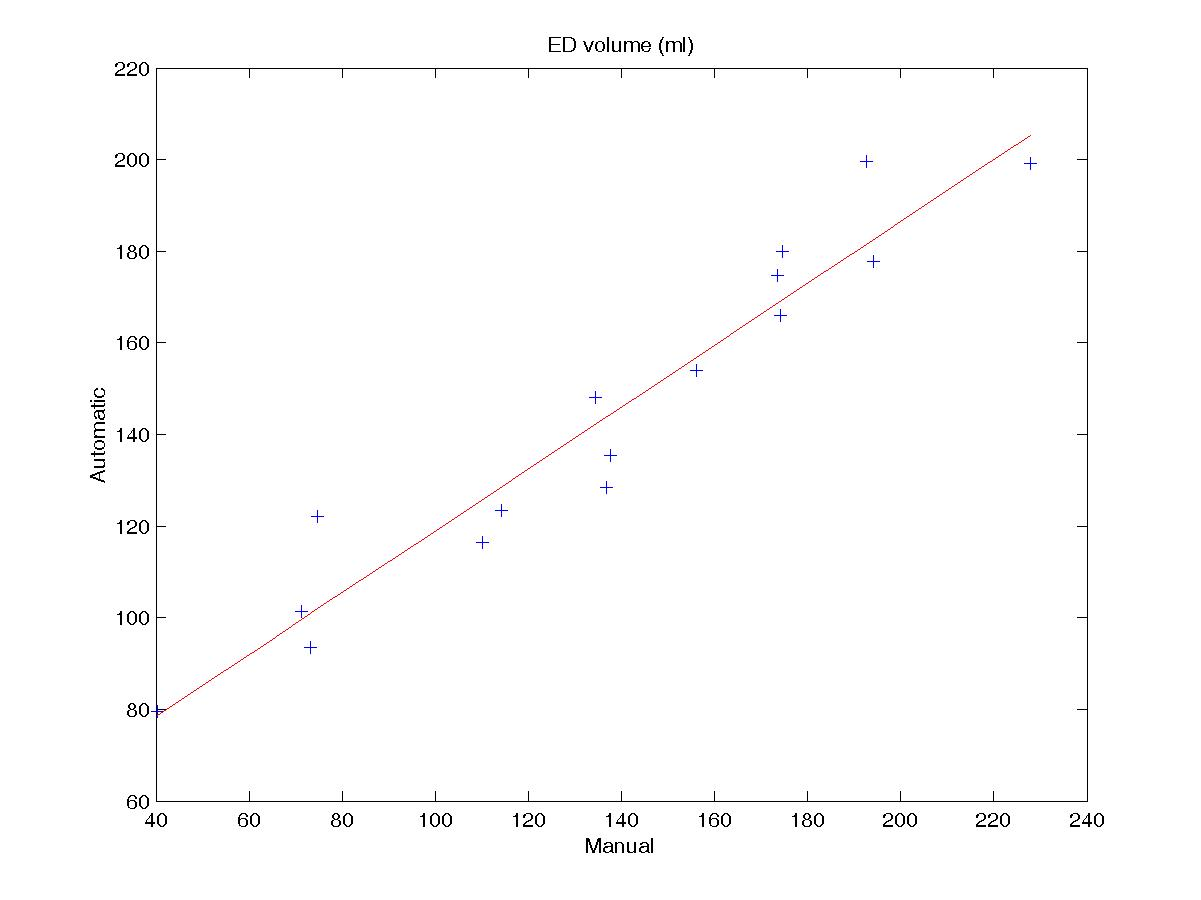} &
\includegraphics[width=0.45\columnwidth]{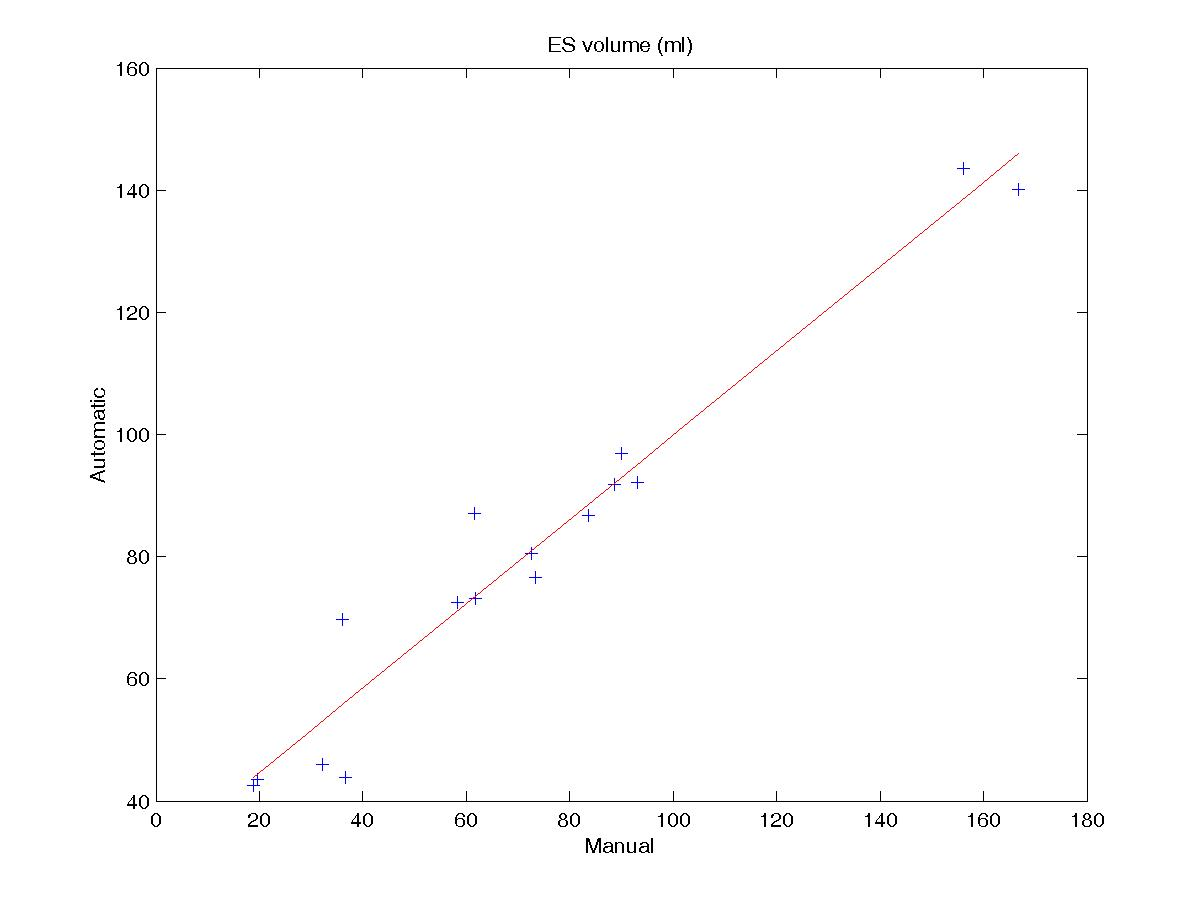}
\end{tabular}
\end{center}
\caption{Regression curve for the endocardial volume at ED (left) and ES (right).  } 
\label{fig:patientresults}
\end{figure}

Figure~\ref{fig:patientresults} shows the regression plots for the endocardial volume at ED and ES.  For both cases, the regression coeffictient is good (0.96 and 0.97 respectively) and the spread of the values is relatively low, demonstrating that the influence of the poor segmentation in the apical slices doesn't have much influence in the endocardial volume estimation. Errors obtained in the computation of the ejection fraction (EF) and ventricular mass (VM), which are based on volume estimations, are reported in table~\ref{tab:resultspatient}. 
  
\section{Discussion and Conclusions}
In this paper, we have presented a multi-atlas propagation and segmentation approach for the segmentation of the RV epicardium and endocardium in ED and ES. As the RV segmentation is done currently in 2D and RV contours exhibit large variation along the heart, we make use of a multi-label ranking criterion, based on the local normalized cross correlation, to select the best atlases for label fusion.  

\begin{table} 
\begin{center}
\caption{Ejection fraction (EF) and ventricular mass (VM) errors.} 
\label{tab:resultspatient} 
\begin{tabular}{lllll} 
\hline\noalign{\smallskip} 
Error & \quad & EF & \quad & VM \\ 
\noalign{\smallskip}
\hline 
\noalign{\smallskip} 
Mean & & 0.15 & & 0.37  \\ 
Standard deviation & & 0.09 & & 0.33 \\ 
\hline 
\end{tabular}
\end{center}
\end{table}

The results show that the method perform well on average but there are some cases in which it fails. In particular, our method has a very high performance on the basal slices, whereas the scores obtained for apical slices are lower (Figure~\ref{fig:segmentations}). These can be explained by two factors: 1) The image quality at the apical slices is rather low. As the registration, the atlas ranking and the label fusion are intensity-based, poor image quality can affect the results, and 2) the regions to be segmented at the apical slices are rather small, which implies that the atlases' masks are also small at these slices. When the masks are used in the registration process to supress undesired structures, the remaining information is insufficient, causing the intensity-based registration, rigid or non-rigid, to fail. 

Nonetheless, these problems can be solved if the registrations are performed in 3D, where more image information can be used for both registration and fusion. Due mainly to slice thickness of this type of images, our current registrations are done in 2D. We will focus our efforts in the extension of the framework so that it can be applied to 3D. 

In conclusion, we have presented a fully automatic segmentation method applied to the extraction of RV epicardium and endocardium. The results show that the method is in average succesful in the segmentation results. The segmentations obtained in the basal slices are of very high quality, whereas the results in the apical ones are less good mainly because the registration stage fails to properly align the atlases with the unseen image. 

\section*{Acknowledgements}
M.A. Zuluaga and S. Ourselin have been founded by the EPRSC Grant 'Grand Challenges: Translating biomedical modelling into the heart of the clinic'   (EP/ H02025X/1). M.J. Cardoso was supported by a scholarship from the Fundacao para a Ciencia e a Tecnologia, Portugal (
SFRH/BD/43894/ 2008). M.J. Cardoso and S. Ourselin receive funding from
the EPSRC grant ' Intelligent Imaging: Motion, Form and Function Across Scale' (EP/H046410/1) and the 
CBRC (Ref. 168).
%
%


\begin{thebibliography}{5}
%
\bibitem {who}
World Health Organization:
Cardiovascular diseases. Fact Sheet No. 317
Online: http://www.who.int/mediacentre/factsheets/fs317/en (2007)

\bibitem {suri}
Suri, J.S..:
Computer vision, pattern recognition and image processing in left ventricle 
segmentation: The last 50 years.
Pattern Anal. Appl. 3(3), 209--242 (2000)

\bibitem {petitjean2011}
Petitjean, C., Dacher J-N.:
A review of segmentation methods in short axis cardiac {MR} images.
Med. Image Anal., 15(2), 169--184 (2011)

\bibitem{caudron}
Caudron J., Fares J., Lefebvre V., Vivier P.H., Petitjean C., Dacher J.N.:
Cardiac MRI assessment of right ventricular function in acquired heart disease: factors of variability.
Acad. Radiol. In press (2012)

\bibitem{zhuang}
Zhuang, X., Rhode, K.S., Razavi, R.S., Hawkes, D.J., Ourselin, S:
A registration-based propagation framework for automatic whole heart segmentation of cardiac {MRI}.
IEEE Trans. Med. Imag. 29(9), 1612--1625 (2010)

\bibitem{warfield}
Warfield, S.K., Zou, K.H., Wells, W.M.:
Simultaneous Truth and Performance Level Estimation ({STAPLE}): An Algorithm for the Validation of Image Segmentation
IEEE Trans. Med. Imag. 23(70, 903--921 (2004)

\bibitem{cardoso}
Cardoso, M.J., Modat, M., Keihaninejad, S., Cash, D., Ourselin, S.:
Multi-STEPS: multi-label similarity and truth estimation for 
propagated segmentations. In:
IEEE Workshop on Mathematical Methods in Biomedical Image Analysis (MMBIA), pp. 153--158 (2012)

\bibitem{ourselin}
Ourselin, S., Roche, S., Subsol, G., Pennec, X., Ayache, N.: 
Reconstructing a 3{D} structure from serial histological sections. 
Image and Vis. Comp., 19(1-2), 25–-31 (2001)

\bibitem{modat}
Modat, M., Ridgway, G.R., Taylor, Z.A., Lehmann, M., Barnes, J., Fox, N.C., Hawkes, D.J., Ourselin, S.:
Fast free-form deformation using graphics processing units.
Comput. Methods Programs Biomed. 98 (3) pp. 278--284(2010)

\end{thebibliography}
\end{document}